\documentclass[superscriptaddress,showpacs,floatfix,twocolumn,prl,reprint]{revtex4-1}

\usepackage{graphicx}
\usepackage{amsmath}

\usepackage{bm}
\usepackage[T1]{fontenc} 
\usepackage{lmodern}
\usepackage{color}
\usepackage{float}

\begin{document}

\title{A universal scaling law for the evolution of granular gases}

\author{Mathias Hummel}
\author{James Clewett}
\author{Marco G. Mazza}
\affiliation{Max Planck Institute for Dynamics and Self-Organisation (MPIDS), Am Fa{\ss}berg 17, 37077 G\"{o}ttingen, Germany}

\date{\today}

\begin{abstract}
Dry, freely evolving granular materials in a dilute gaseous state coalesce into dense clusters only due to dissipative interactions. This clustering transition is important for a number of problems ranging from geophysics to cosmology. Here we show that the evolution of a dilute, freely cooling granular gas is determined in a universal way by the ratio of inertial flow and thermal velocities, that is, the Mach number. Theoretical calculations and direct numerical simulations of the granular Navier--Stokes equations show that irrespective of the coefficient of restitution, density or initial velocity distribution, the density fluctuations follow a universal quadratic dependence on the system's Mach number. We find that the clustering exhibits a scale-free dynamics but the clustered state becomes observable when the Mach number is approximately of $\mathcal{O}(1)$. Our results provide a method to determine the age of a granular gas and predict the macroscopic appearance of clusters.
\end{abstract}
   
\maketitle

\begin{figure}[htpb]
\centering
    \includegraphics[height = 0.48\columnwidth ,width = 0.48\columnwidth]{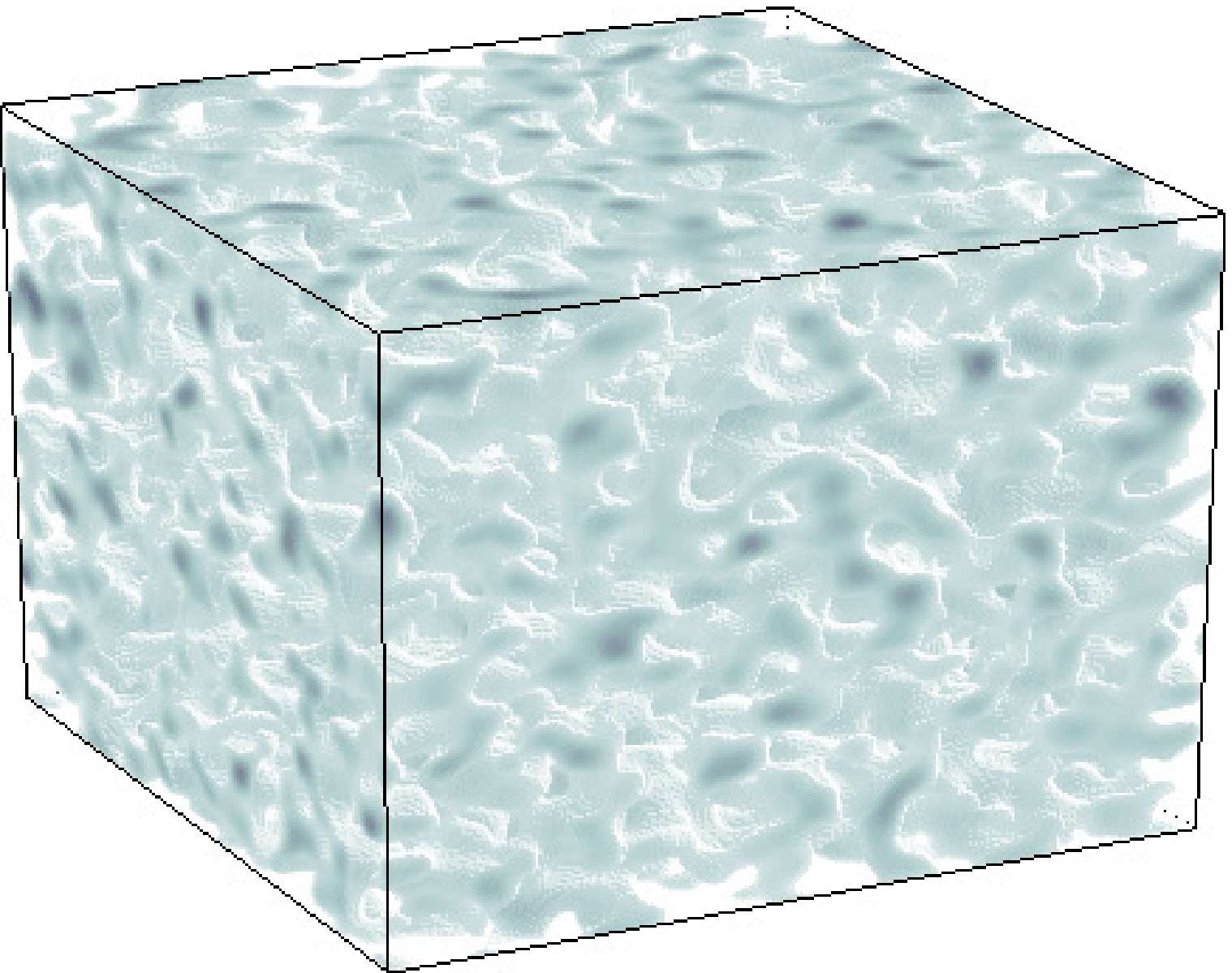}
    \includegraphics[height = 0.48\columnwidth ,width = 0.48\columnwidth]{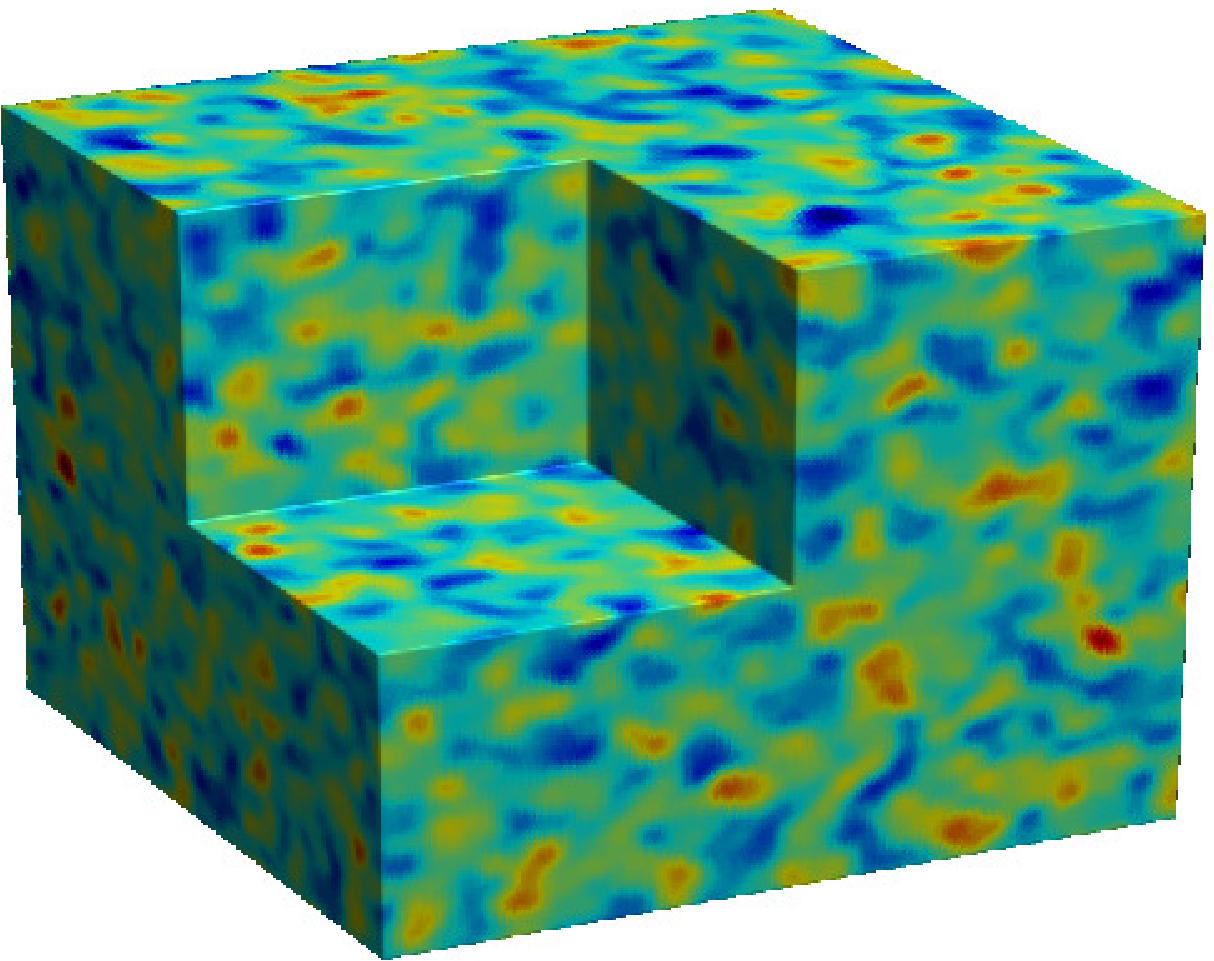}
    \includegraphics[height = 0.48\columnwidth ,width = 0.48\columnwidth]{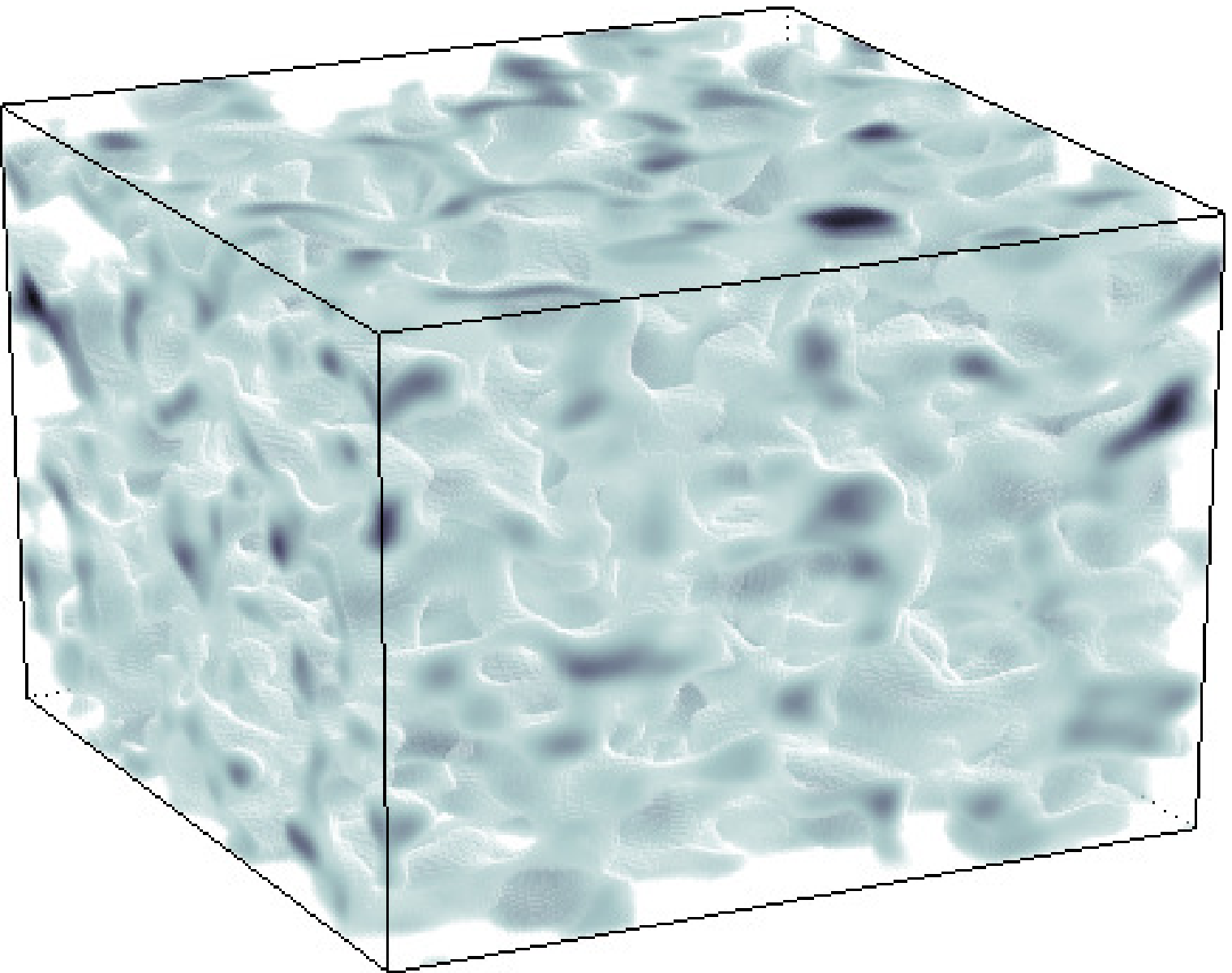}
    \includegraphics[height = 0.48\columnwidth ,width = 0.48\columnwidth]{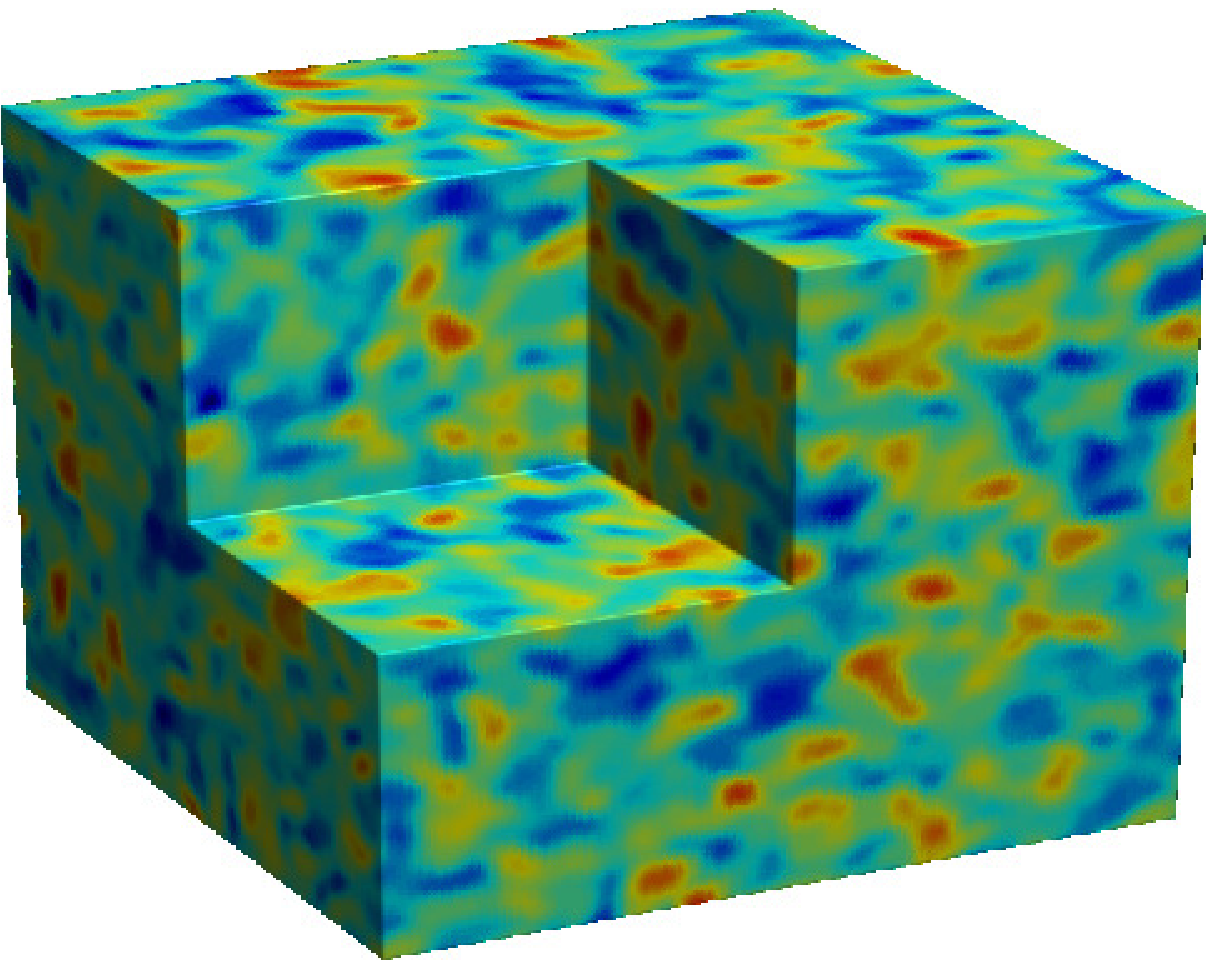}
    \includegraphics[height = 0.48\columnwidth ,width = 0.48\columnwidth]{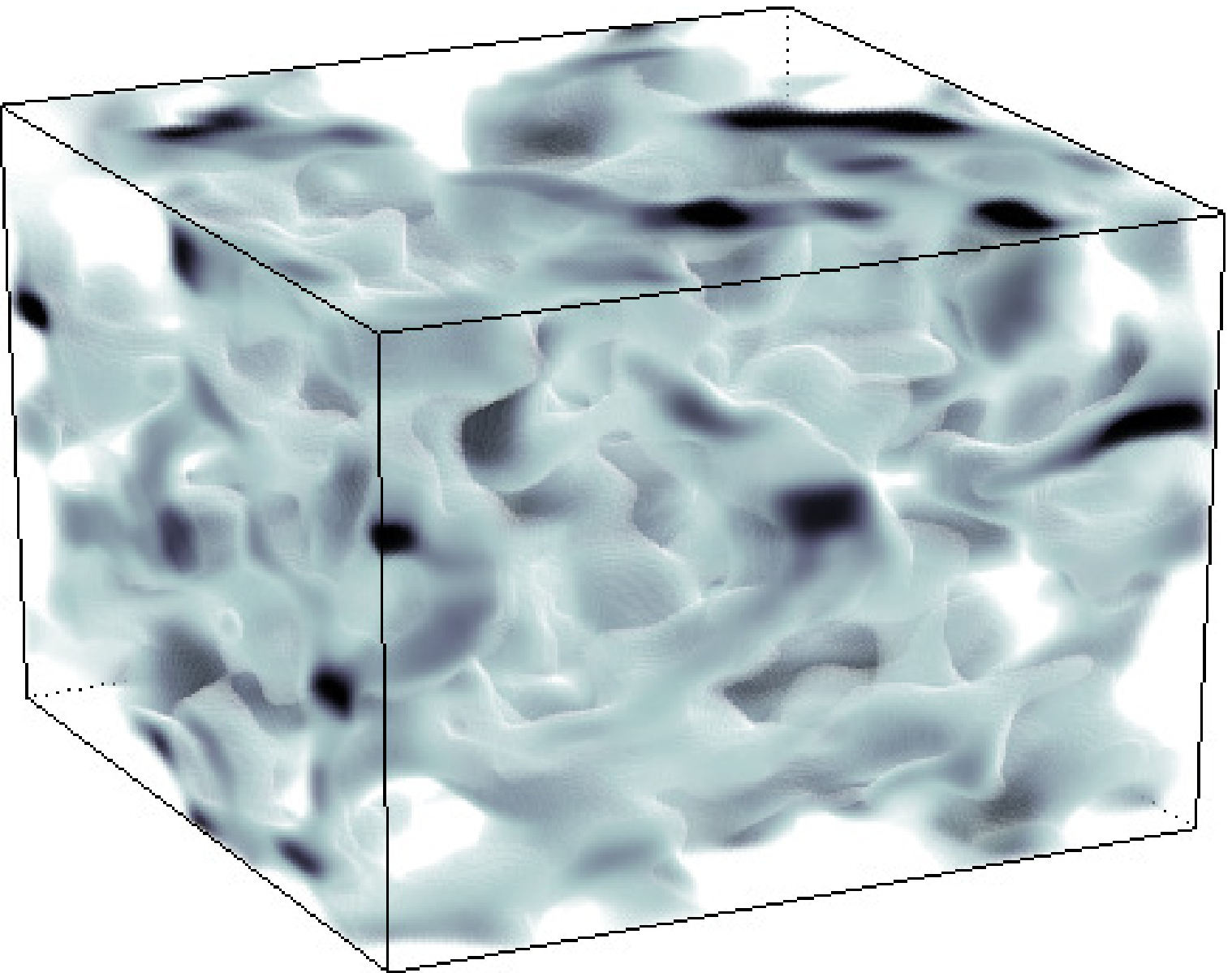}
    \includegraphics[height = 0.48\columnwidth ,width = 0.48\columnwidth]{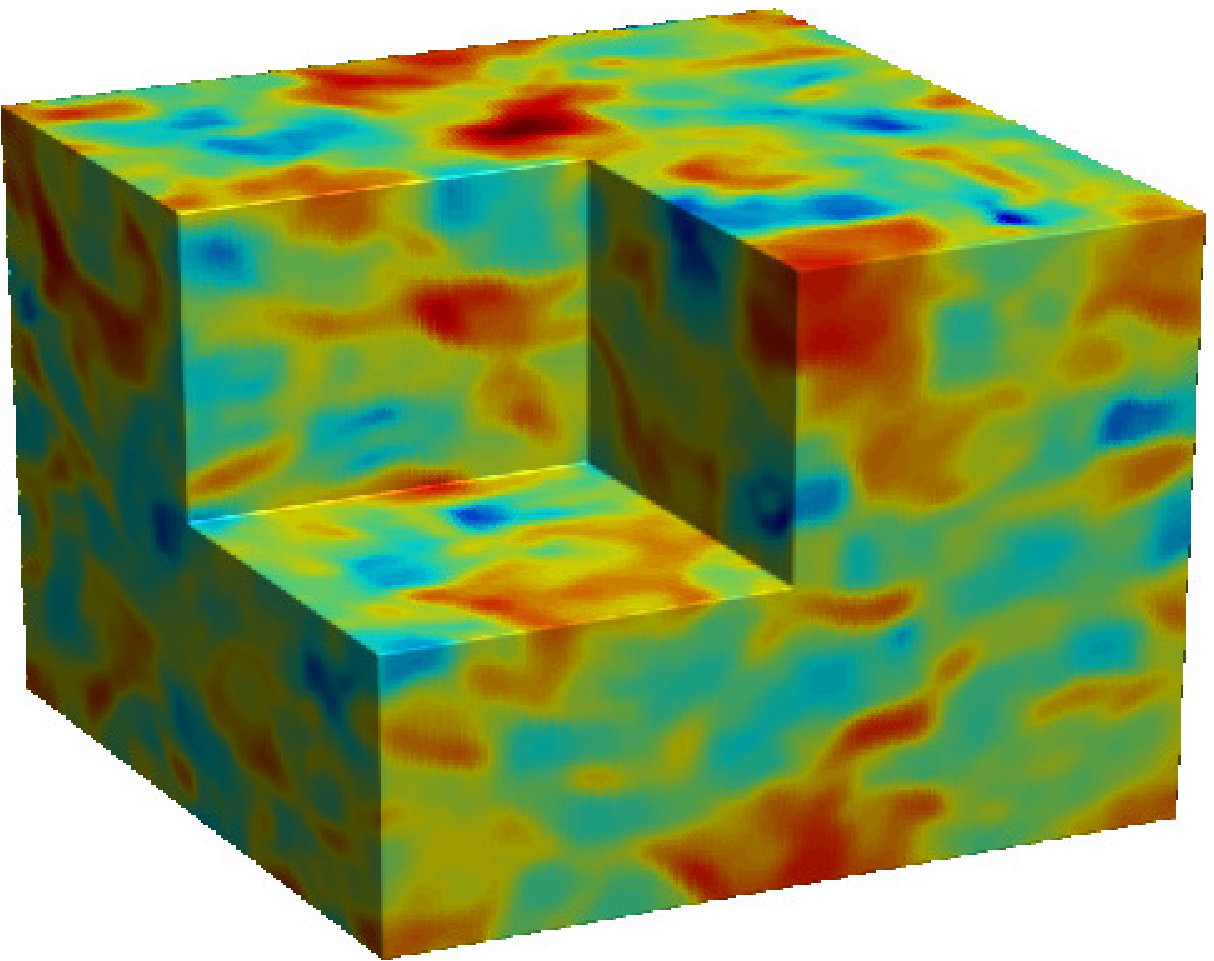}
\caption{Snapshots of the system showing the temporal evolution of a three-dimensional granular gas with a coefficient of restitution $\varepsilon = 0.9$ and average filling fraction $\bar{\phi}=0.05$.
    (Left column) From top to bottom snapshots of the system density at times $t= 10^9$, $t= 10^{10}$ and $t= 10^{11}$, respectively.
    The grey scale represent the local average density.
    (Right column) Three-dimensional map of the temperature field $T(\vec{r},t)$ at the same times to the corresponding density plot in the left column.
    The colour (red - hot; blue - cold) represents the local temperature compared to the average temperature in the system.
}
\label{fig:snap_shots}
\end{figure}

Geophysical processes~\cite{goldhirschARFM2003}, the solar corona~\cite{fieldAJ1965}, the asteroid belt between Mars and Jupiter, planetary rings~\cite{goldreichARAA1982,bridgesN1984}, protoplanetary disks~\cite{johansenAPJ2007}, and the formation of cosmological structures~\cite{shandarinRMP1989} are systems where granular clustering processes are at play. 
Even a small degree of dissipation in the kinetics of granular particles produces spatial correlations and structures in a dilute, homogeneous gas~\cite{GoldhirschPRL1993}.
 Hydrodynamic treatments suggest that a shear instability initiates this transition~\cite{GoldhirschPRL1993}. However, what exactly initiates this process is not known. 
The equations of granular hydrodynamics~\cite{BrilliantovBook2004} predict a linear instability of the transverse mode~\cite{McNamara1993} when the wavevector $\|\vec k\|\le k_\perp^*(\varepsilon)$, where $k_\perp^*(\varepsilon)\propto(1-\varepsilon^2)^{1/2}$ and $\varepsilon$ is the coefficient of restitution. This instability leads to the formation of vortices. In regions where the particle velocities are correlated the temperature drops, which, in turn, creates a region of low pressure. These are the seeds for a second instability if the system size is larger than $k_\mathrm{H}^{*-1}(\varepsilon)>k_\perp^{*-1}(\varepsilon)$. Although freely cooling granular gases have attracted wide interest~\cite{deltourJdP1997,ludingChaos1999,MillerPRE2004,efratiPRL2005,meersonPRE2008,
puglisiPRE2008,kolvinPRE2010} there remains, beyond issues of finite sizes, the outstanding question of what sets the rise of the density inhomogeneities. Past works made differing claims as to what the onset time of clustering is~\cite{puglisiPRE2008,ahmadPRE2007,efratiPRL2005}. 

Experimental studies~\cite{falconPRL1999,MaassPRL2008,heisselmannIcarus2010,sackPRL2013,harthASR2015} of granular gases are rather scarce because of formidable challenges in preparing a system free of external forces. Microgravity experiments on parabolic flights or drop towers potentially offer good conditions, but it is very difficult to remove the influence of confining potentials or surrounding walls on the granular system.
The necessity of studying large system sizes and of characterizing fluctuations in regions of sharp gradients  in temperature and density, developing into supersonic flow, without ambiguities motivates us to tackle the continuous hydrodynamic equations beyond perturbative schemes.  
Hydrodynamic fields can be rigorously defined in a manner similar to molecular fluids by means of a coarse-graining of the microscopic kinetic equations. First, one considers the one-particle distribution function $f(\vec r, \vec w,t)$, which obeys the Boltzmann equation and represents the number of particles within a volume $\mathrm{d}\vec{r}$ centered at $\vec{r}$ and with velocity $\vec{w}$ within the interval $\mathrm{d}\vec{w}$. Then, the transport equations for inelastic systems are derived by using a Chapman--Enskog expansion~\cite{breyJSP1997,breyPRE1998,garzoPRE1999}. From the moments of $f(\vec r, \vec w,t)$ the density $\rho(\vec r,t)$, flow velocity $\vec{v}(\vec r,t)$, and  temperature $T(\vec r,t)$ fields can be derived. Physically, $T(\vec r,t)$ represents the fluctuation of the microscopic velocities $\vec w$.

Direct numerical simulations (DNS) of the hydrodynamic equations  give access to enlightening results because all hydrodynamic fields are accessible at each point of space and time. Therefore, DNS allows for the possibility of observing fluctuations and structure formation on scales inaccessible to molecular dynamics simulations.
\begin{figure}
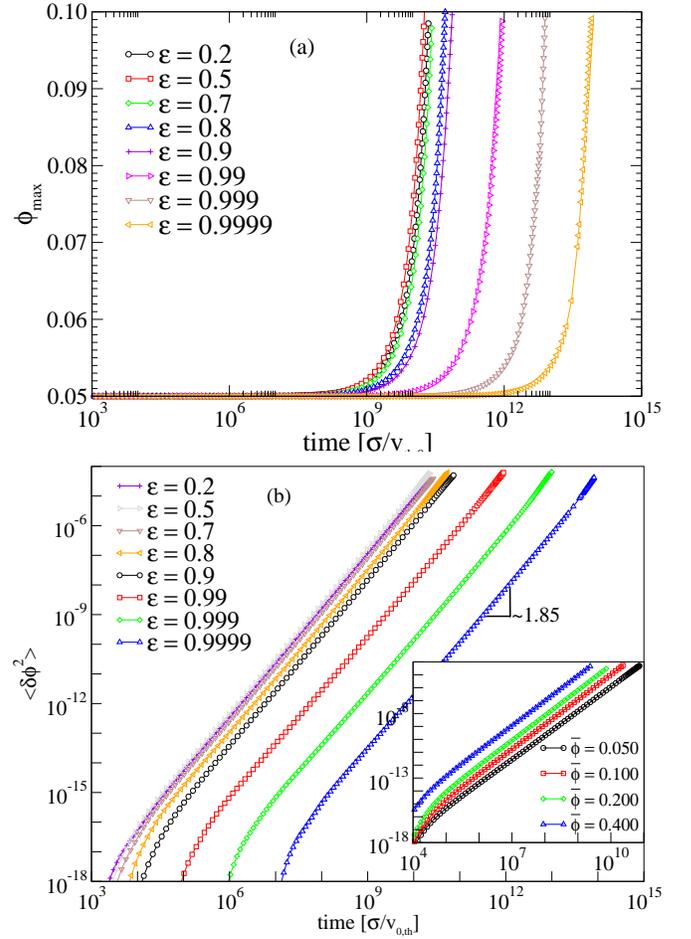

\centering
\includegraphics[width = 1\columnwidth]{epsilon_maxPhiTime}
\includegraphics[width = 1\columnwidth]{Time_EpsilonANDFilling}
\caption{Temporal evolution of the maximum filling fraction ${\phi}_\mathrm{max}$ (a) and of the density fluctuations $\langle\delta \phi^2\rangle$ (b).
The clustering process exhibits the same qualitative features over a wide range of coefficients of restitution $\varepsilon$ and average filling fractions $\overline{\phi}$. 
The curves are guides to the eye.
As $\varepsilon$ grows, the time of onset of clustering increases of four orders of magnitude.
The inset panel shows the effect of varying average filling fractions.
  }
\label{fig:maxdensity}
\end{figure}
\begin{figure}
\centering
\includegraphics[width = 1\columnwidth]{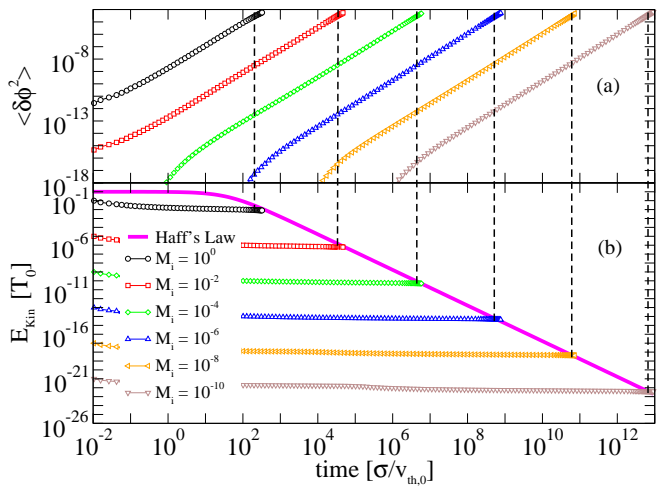}
  \caption{(a) Temporal evolution of density fluctuations for different initial values of the Mach number $\mathcal{M}$.
  (b) Evolution of the kinetic energy (symbols) for the same initial Mach numbers as in (a), and evolution of the temperature (solid, magenta line). 
  We mark the times $t^*(\mathcal{M}_i)$ when the kinetic energy equals the temperature and the corresponding values of $\langle\delta\phi^2\rangle(t^*)$ (vertical, dotted lines). 
  To test the hypothesis of an independent behaviour in terms of $\mathcal{M}$, we vary the initial, inertial velocities in the system, that is, the initial value of the Mach number $\mathcal{M}_i$. 
  We find that for larger $\mathcal{M}_i$ the onset of clustering occurs decades in time earlier, but still it coincides with the time when $\mathcal{M}\sim O(1)$.
}
\label{fig:phiEnergyTime}
\end{figure}
We solve the compressible Navier--Stokes equations for $\rho$, momentum $m\vec v$, and $T$ of granular matter in three dimensions~\cite{BrilliantovBook2004}.
The heat flux  for granular materials contains a ``pycnothermal'' term in addition to thermal gradients, that is $\vec q = -\kappa \nabla T - \mu \nabla \rho$~\cite{sotoPRL1999}. 
The term proportional to the density gradient is in principle present also in molecular fluids but the Onsager theorem protects against it~\cite{duftyJPCC2007} yielding $\mu=0$. 
In contrast, for granular gases the coarse-graining of particles' degrees of freedom that generates hydrodynamic fields produces a non-vanishing $\mu$~\cite{candelaAJP2007}. 
We also include the Carnahan--Starling pair-correlation function to realistically represent spatial correlations. 

\begin{figure*}
\centering
\includegraphics[width = 1.5\columnwidth]{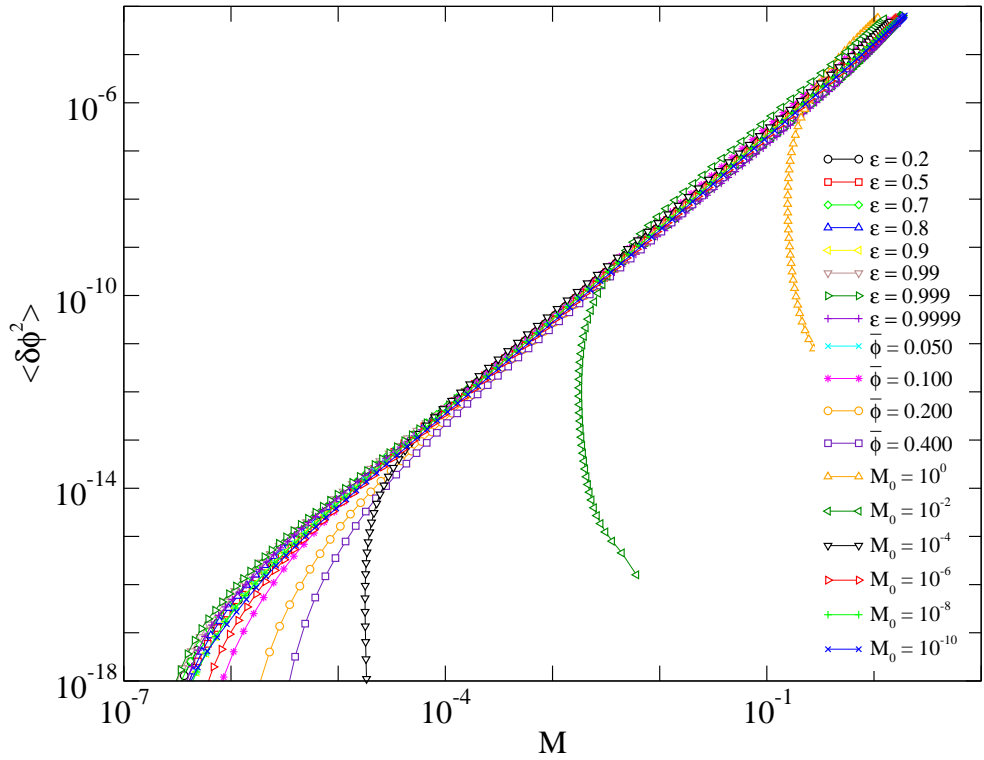}
\caption{The density fluctuations exhibit a universal scaling in terms of the system's average  Mach number $\mathcal{M}$.   After the relaxation of the initial conditions the granular gas shows a collapse of the density fluctuations on the curve $\langle\delta\rho^2\rangle(\mathcal{M})=c\mathcal{M}^2$. Calculations are shown for system at $\bar\phi=0.05$, $\varepsilon=0.9$ and $\mathcal{M}_0=10^{-8}$ unless the variable are explicitly changed according to the legend. 
}
\label{fig:scaling}
\end{figure*}
Figure \ref{fig:snap_shots} shows the evolution of $\rho(\vec r,t)$ and $T(\vec r,t)$ in a freely cooling gas.
We observe that out of the homogeneous cooling state small, cold regions of larger density emerge throughout the system and grow in size with time. The high density regions (clusters) are filamentary because of the shear instability. Similar structures have been seen in molecular dynamics simulations~\cite{ludingPRAMANA2005,pathakPRL2014}, however our system size allows the formation of multiple, large clusters. 
We quantify the cluster growth by  calculating the maximum filling fraction $\phi_{max}$ and the density fluctuations $\langle\delta \rho^2\rangle$ across the system, which allows one to compare systems with different average filling fractions $\bar\phi$. Figure~\ref{fig:maxdensity} shows the temporal evolution of these quantities.  
In the homogeneous cooling state $\phi_{max}$ is very small, but suddenly the systems develops inhomogeneities. Figure~\ref{fig:maxdensity}(b) reveals an early time dynamics not directly accessible when observing the maximum or the average cluster size (Fig. \ref{fig:maxdensity}(a)). 
What appears to be a sudden emergence of clusters at characteristic times depending on $\varepsilon$ in Fig.\ref{fig:maxdensity}(a) is instead a scale-free process described by a power law $\sim t^{\alpha}$, with $\alpha=1.85$. Thus, if the signal-to-noise ratio of the measurements is not large enough an apparent sudden onset of clustering will be visible only when $\phi_{max}$ has reached few percents. 
To understand the origin of the increase of the density we examined the behaviour of the average kinetic energy $K(t)\equiv\sum_i\tfrac{1}{2}mv_i^2$ and temperature $T(t)$ and compared it with the temporal evolution of $\phi_{max}$. 
The temperature follows Haff's law $T(t) = T_0\left(1+t/\tau\right)^{-2},\tau \propto \overline{\phi}\left(1-\varepsilon^{2}\right)\sqrt{T_0} $ while $K(t)$ decays with time in an much weaker way. 
Figure~\ref{fig:phiEnergyTime} shows that the size of inhomogeneities is directly linked to the relative decrease of the kinetic energy $K(t)$ with respect to the temperature. In fact, when the kinetic energy equals the temperature (Haff's law), $\langle \delta \phi^2 \rangle$ reaches the same value $\approx 10^{-4}$ for all our simulations. 
This point coincides with the time when clusters would become visible in an experiment or molecular dynamic simulation (i.e. $\Delta\phi\sim\sqrt{\langle\delta\phi^2\rangle}\approx 1\%$).
Therefore, it seems appropriate to us to consider the Mach number $\mathcal{M} \equiv \sqrt{\langle v^2\rangle/\langle T \rangle}$, as an invariant measure of the granular dynamics, where the angle brackets indicate spatial averages. 
The granular gas develops visible clusters when $\mathcal{M}$ is of the order of one, that is the threshold of supersonic flow.
 Changing the average filling fraction or coefficient of restitution does not alter this conclusion.

Because the system is translational invariant and there is no characteristic time scale associated with the condition $\mathcal{M}\sim \mathcal{O}(1)$ we expect the evolution of the freely cooling gas to be scale invariant when described with the relevant variables. 
In fact, under general assumptions we can prove with an analytical argument that the density fluctuations $\langle\delta\rho^2\rangle$ scale quadratically with $\mathcal{M}$.
We assume that: (i)  the granular gas is in a homogeneous cooling state where Haff's law holds,  $T = T_0\left(1+t/ \tau\right)^{-2}$, and the density fluctuations are still small, that is, $\rho = \rho_0 +\delta\rho$, $\delta\rho \ll \rho_0$; 
(ii)  isotropy of the velocity, $\langle v_x\rangle=\langle v_y\rangle=\langle v_z\rangle$;
(iii) the equation of state of ideal gases holds: $p = \rho T$; 
(iv)   the system is so large that we can neglect finite size effects and that diffusion is negligible; 
(v) local fluctuations in the bulk velocities are small: $v \ll T^{1/2}$.
Then the granular Navier--Stokes equations  simplify to
\begin{gather}
\begin{aligned}\label{eq:approxNS}
    \partial_t \rho + \nabla \cdot (\rho \vec v) = 0\,~\\
    \partial_t \vec v +\nabla(\rho T) = 0\,.
  \end{aligned}
\end{gather}
From taking the spatial averages of Eqs.~\ref{eq:approxNS} it follows that $\langle\delta\rho^2\rangle\sim\mathcal{M}^2$
(see the supplementary information for more details).
Figure~\ref{fig:scaling} collects results from extensive DNS calculations where we varied $\varepsilon$, $\mathcal{M}_0$, $\bar{\phi}$ and shows the evolution of $\langle\delta\rho^2\rangle$ 
in terms $\mathcal{M}$. 
Regardless of the system parameters or the initial state, the density fluctuations converge onto the locus $\langle\delta\rho^2\rangle(\mathcal{M})=c\mathcal{M}^2$, where $c$ is a constant (the effect of varying initial conditions is described in the supplementary information).
This locus plays the role of an ``attractor'' for the evolution of the granular gas. 
Once the initial conditions are forgotten, all systems investigated show universal behaviour, as visible from the collapse of all curves in Fig.~\ref{fig:scaling} over several decades.
In summary, the universal power-law behaviour of a granular gas in the homogeneous cooling state indicates that there is no characteristic time scale for the onset of clustering. However, because of finite resolution, an experiment would observe the onset of clustering when $\mathcal{M}\sim\mathcal{O}(1)$.

We gratefully acknowledge K. Binder and S. Herminghaus for helpful conversations and for critically reading this manuscript. 

%

\end{document}